\documentclass[sn-mathphys-num]{sn-jnl}

\usepackage{graphicx}%
\usepackage{multirow}%
\usepackage{amsmath,amssymb,amsfonts}%
\usepackage{amsthm}%
\usepackage{mathrsfs}%
\usepackage[title]{appendix}%
\usepackage{xcolor}%
\usepackage{textcomp}%
\usepackage{manyfoot}%
\usepackage{booktabs}%
\usepackage{algorithm}%
\usepackage{algorithmicx}%
\usepackage{algpseudocode}%
\usepackage{listings}%
\usepackage{mathtools}
\usepackage{amsfonts}
\usepackage{amsmath}
\usepackage{amsthm}
\usepackage{nomencl}
\usepackage{courier}
\usepackage{array}
\usepackage{tabularx}
\usepackage{multirow}
\usepackage{fancyvrb}
\usepackage{booktabs}
\usepackage{makecell}
\usepackage{placeins}
\usepackage{enumitem}
\usepackage{amsmath}
\usepackage{braket}
\usepackage[utf8]{inputenc}
\usepackage[T1]{fontenc}

\begin{document}

\title[Article Title]{Implementation and Analysis of Regev’s Quantum Factorization Algorithm}

\author{\fnm{Przemysław} \sur{Pawlitko}}\email{przemyslaw.pawlitko@tele.agh.edu.pl}

\author{\fnm{Natalia} \sur{Moćko}}\email{natalia.mocko@tele.agh.edu.pl}

\author{\fnm{Marcin} \sur{Niemiec}}\email{marcin.niemiec@agh.edu.pl}

\author*{\fnm{Piotr} \sur{Chołda}}\email{piotr.cholda@agh.edu.pl}

\affil{\orgname{AGH University of Krakow}, \orgaddress{\street{Mickiewicza 30}, \city{Krakow}, \postcode{30-059}, \country{Poland}}}

\abstract{
Quantum computing represents a significant advancement in computational capabilities. 
Of particular concern is its impact on asymmetric cryptography through, notably, Shor's algorithm and the more recently developed Regev's algorithm for factoring composite numbers. We present our implementation of the latter. 
Our analysis encompasses both quantum simulation results and classical component examples, with particular emphasis on comparative cases between Regev's and Shor's algorithms. 
Our experimental results reveal that Regev's algorithm indeed outperforms Shor's algorithm for certain composite numbers in practice. However, we observed significant performance variations across different input values. Despite Regev's algorithm's theoretical asymptotic efficiency advantage, our implementation exhibited execution times longer than Shor's algorithm for small integer factorization in both quantum and classical components. These findings offer insights into the practical challenges and performance characteristics of implementing Regev's algorithm in realistic quantum computing scenarios.
}



\maketitle

\section{Introduction}\label{sec1}

Public-key (asymmetric) cryptography is a fundamental pillar of modern cybersecurity. Its principles underpin the security of numerous systems, including secure Internet communication, authentication protocols, banking transactions, cloud data protection, software integrity verification, medical data encryption, and many other applications. Without secure public-key cryptography, sensitive information would be exposed, privacy compromised, and the risk of online fraud and cybercrime would increase significantly. One of the most widely used public key algorithms, RSA~\cite{RSA1978}, is based on the mathematical difficulty of factorizing a large semiprime integer $N$, which is the product of two prime numbers $p$ and $q$. For classical computers, factorizing such a number is computationally infeasible within a practical, i.e., polynomial, timeframe. For example, breaking RSA encryption with $2048$-bit key and by computing $1.6 \times 10^{16}$ operations per second would require approximately $19.8$ quadrillion years using the brute-force method and classical computational techniques. Moreover, the compromise of RSA would have far-reaching implications for other cryptographic algorithms that are based on similar computational problem principles (i.e., those based on discrete logarithms).

Although breaking RSA encryption is currently impossible for classical computers, in 1994 the American computer scientist and mathematician Peter Shor published a work ``\textit{Polynomial-Time Algorithms for Prime Factorization and Discrete Logarithms on a Quantum Computer}''~\cite{shor1994, shor1999} that proved that factorizing large numbers efficiently is possible using quantum computers. His algorithm had profound implications for cryptography, raising awareness that current public-key encryption methods would not remain secure in the era of advanced quantum computing.

However, the development of scalable and powerful quantum computers capable of running Shor's algorithm remains a significant technical challenge. As the number of qubits and quantum gates increases, quantum computers become more susceptible to quantum decoherence. Decoherence disrupts quantum effects, which are essential for the proper execution of quantum algorithms. Shor's algorithm, in particular, requires a large number of quantum gates, making it desirable to minimize this requirement to improve the feasibility of practical implementations.

To address these challenges, Oded Regev proposed an alternative quantum algorithm in his recent work, ``\textit{An Efficient Quantum Factoring Algorithm}''~\cite{regev2023v1, regev2023v2, regev2024, Regev2025}\footnote{Regev published four versions of this algorithm, with the first appearing in August 2023, and the most recent one in January 2025. This paper references the latest version.}, which can be thought of as a multidimensional extension of Shor's algorithm. Regev's algorithm reduces the number of quantum gates required for factorization compared to Shor's algorithm. However, Regev acknowledges that it remains unclear whether this reduction will translate into practical improvements in the physical implementation of quantum computers. Despite this uncertainty, his work represents a critical step toward overcoming the obstacles associated with quantum computing and advancing the field of quantum cryptography. 

\subsection{Contributions}

At the time of working on this paper, no publicly available implementation of Regev's quantum algorithm for quantum computers existed. Therefore, the first challenge was to develop the first implementation of Regev's algorithm for quantum computers. The implementation served as the foundation for a detailed analysis of its speed and efficiency, achieved by introducing and varying relevant parameters. Throughout our study, comparisons were made between Regev's algorithm and Shor's algorithm to highlight their respective strengths and limitations. Additionally, this paper aimed to identify and present potential areas for future research on Regev's algorithm, providing a starting point for subsequent studies in the field.

\subsection{Structure of the Paper}

The remainder of this paper is organized as follows. First, in Section~\ref{sec2}, both Shor's and Regev's algorithms are reminded and shorty compared, especially from the viewpoint of feasible implementations of the quantum part. The next part, Section~\ref{cha:regev_algorithm_implementation} focuses on the non-trivial details of the Regev's algorithm implementation. Section~\ref{an1} elaborates on the comparison of both Regev's and Shor's algorithms from the viewpoint of performance, although the results related to the former are emphasized as more original. Finally, Section~\ref{concl} concludes the paper.

\section{Basics of Shor's and Regev's Algorithms}\label{sec2}

Both algorithms aim to factorize a semiprime $N$, an $n$-bit integer that is the product of prime numbers $p,q \in \mathbb{P}$ (i.e., $N=p \cdot q$). However, the factorization process is carried out in different ways.

\subsection{Idea of the Algorithms}

Basically, Shor's algorithm finds the smallest even period $r$ of a modular exponentiation function, which is defined as follows:

\[\quad z \mapsto a^z \bmod{N}\]

Here, $a$ is a random integer that belongs to the group of units of $(\mathbb{Z}/N\mathbb{Z})^\times$ and $z \in \mathbb{Z}$. After finding the period of this function, factorizing $N$ is trivial and begins with this equation:

\[a^r \equiv 1 \bmod{N}\]

During the next steps, we use the condition that $r$ is an even number:

\[a^r - 1 \equiv 0 \bmod{N}\]
\[\left(a^{\frac{r}{2}} - 1\right)\left(a^{\frac{r}{2}} + 1\right) \equiv 0 \bmod{N}\]
\[\left(a^{\frac{r}{2}} - 1\right)\left(a^{\frac{r}{2}} + 1\right) \equiv 0 \bmod{(p\cdot q)}\]

Therefore, there must exist such $k \in \mathbb{Z}$, that:

\[\left(a^{\frac{r}{2}} - 1\right)\left(a^{\frac{r}{2}} + 1\right) = p\cdot q\cdot k\]

As a result, it is very likely that $p$ is the divisor of $\left(a^{\frac{r}{2}} - 1\right)$ and $q$ is the divisor of $\left(a^{\frac{r}{2}} + 1\right)$\footnote{As discussed by Shor~\cite{shor1999}, this probability is ``at least $1-1/2^{k-1}$, where $k$ is the number of distinct odd prime factors of $n$'', so in this case approximately 50\% chance. One possibility is that both $p$ and $q$ are factors of $(a^{\frac{r}{2}} + 1)$ or $(a^{\frac{r}{2}} - 1)$. Theoretically, there should not be any case that they are both factors of $(a^{\frac{r}{2}} - 1)$, but sometimes Shor's algorithm finds a multiplication of the smallest period.}:

\[p = \gcd\left(\left(a^{\frac{r}{2}} - 1\right), N\right)\]
\[q = \gcd\left(\left(a^{\frac{r}{2}} + 1\right), N\right) = \frac{N}{p}\]

If it comes to the implementation, the most time-consuming part of the algorithm is a previously described modular exponentiation function:

\[\quad z \mapsto a^z \bmod{N}\]

The reason is that seeing the period of this function requires $z$ to be large enough. For the $n$-bit integer $N$, $z$ increases to $2^n$. Moreover, this function needs to be applied in superposition inside the quantum part of the algorithm. In order to reduce a computational load, the \textit{repeated squaring trick} is used. Thus, quantum part needs $n$ multiplications of $n$-bit numbers, so roughly $\widetilde{O}(n^2)$ gates\footnote{The Big-$O$-tilde notation ($\widetilde{O}$) is commonly used in describing quantum metrics. Unlike the classical Big-$O$ notation, it disregards logarithmic factors. For example, $O(n\log{n}) = \widetilde{O}(n)$.}. But still, the numbers that need to be computed are very large.

There is also an even more serious problem. Quantum effects, such as superposition and entanglement, can be destroyed during the run of an algorithm. This is because the larger the number of entangled particles and~/~or the larger the number of gates in a quantum circuit, the more quantum effects are susceptible to noise and decoherence, leading to destruction~\cite{mutter2023}. For this reason, it is desirable to reduce the number of quantum gates as much as possible.

To address these challenges, Oded Regev proposed the approach of extending to a larger number of dimensions~\cite{regev2024}. The concept of his algorithm bears similarities to Shor's algorithm, as it also focuses on finding the period of a specific function. However, the function is defined in a slightly different manner:

\begin{equation}
(z_1, z_2, \ldots, z_d) \mapsto a_1^{z_1}\cdot a_2^{z_2}\cdot \ldots \cdot a_d^{z_d} \bmod{N}
\label{eq:regev_periodic_function}
\end{equation}

Here, the numbers $a_1, a_2, ..., a_d$ are the first squared primes (that is, $4, 9, 25, ...$), $N$ is still an integer of $n$ bits that is being factorized, and $z_i \in \mathbb{Z}$. After finding the period vector $[r_1, r_2, ..., r_d]$ of this function, factorization of $N$ is trivial, and -- similarly to the transformations described above -- it starts with the following step:

\[a_1^{r_1}\cdot a_2^{r_2}\cdot \ ...\ \cdot a_d^{r_d} \equiv 1 \bmod{N}\]

Let us define $a_i = b_i^2$. Due to the fact that each $a_i$ is a squared prime $b_i$:

\[b_1^{2^{r_1}}\cdot b_2^{2^{r_2}}\cdot \ ...\ \cdot b_d^{2^{r_d}} \equiv 1 \bmod{N}\]
\[b_1^{2^{r_1}}\cdot b_2^{2^{r_2}}\cdot \ ...\ \cdot b_d^{2^{r_d}} - 1 \equiv 0 \bmod{N}\]
\[\left(b_1^{r_1}\cdot b_2^{r_2}\cdot \ ...\ \cdot b_d^{r_d} - 1\right)\left(b_1^{r_1}\cdot b_2^{r_2}\cdot \ ...\ \cdot b_d^{r_d} + 1\right) \equiv 0 \bmod{N}\]


Regev's algorithm relies on a number-theoretic heuristic assumption reminiscent, as described in~\cite{regev2024}. This assumption states that at least half of the period vectors, when applied to the analyzed cyclic function and followed by bilateral square rooting, produce non-trivial factors of $1 \bmod{N}$, that is, factors that are neither $1$ nor $-1$. Consequently, this assumption ensures that, in at least half of the cases, $b_1^{r_1}\cdot b_2^{r_2}\cdot ... \cdot b_d^{r_d} \notin \{1, -1\} \bmod{N}$. As a result, it avoids the trivial solutions $\left(b_1^{r_1}\cdot b_2^{r_2}\cdot ... \cdot b_d^{r_d} - 1\right) = 0$ or $\left(b_1^{r_1}\cdot b_2^{r_2}\cdot ... \cdot b_d^{r_d} + 1\right) = 0$, thereby enabling the computation of $p$ and $q$, as detailed in the following discussion.

From this point, the reasoning aligns with that of Shor's algorithm:

\[p = \gcd\left(\left(b_1^{r_1}\cdot b_2^{r_2}\cdot \ ...\ \cdot b_d^{r_d} - 1\right), N\right)\]
\[q = \gcd\left(\left(b_1^{r_1}\cdot b_2^{r_2}\cdot \ ...\ \cdot b_d^{r_d} + 1\right), N\right) = \frac{N}{p}\]

In the next lines, we present a sketch of the proof that this approach will also result in a period vector finding. It is based on the brief proof description that Regev outlined in his work.

To find a period, there need to be at least two combinations of the vectors $(z_1, z_2, ...,z_d)$ that give the same function results. Following this condition, a period vector can be constructed by analyzing these two vector values.

The analyzed function~\eqref{eq:regev_periodic_function} is a product of $d$ squared primes raised to a power of up to $2^{n/d}$. Thus, the total number of results of the analyzed function is $(2^{n/d})^d = 2^n$. 

Furthermore, period $r$ of the analyzed function abides by this inequality:

\[r \leq \phi(N) = (p-1)(q-1) = pq - p - q + 1 < N = pq \leq 2^n-1 < 2^n\]

Combining these two facts and using the pigeon hole principle, there have to be two vectors $(z_1, z_2, ...,z_d)$ giving the same function result and that is why the period can be found.

With this approach, each $z_i$ goes up to $2^{n/d}$ and not up to $2^n$ as in Shor's approach, which reduces the scale of the computed numbers, and thus reduces the computational load. Moreover, it appears that the number of gates needed in the quantum part of the original Regev's algorithm is reduced to $\widetilde{O}\left(n^{3/2}\right)$, which reduces the noise and decoherence of quantum effects. In our case, the gate complexity is on the order of $\widetilde{O}\left(n^{5/2}\log n\right)$, whereas the width of the quantum circuit is smaller, on the order of $\widetilde{O}(n)$. 

\subsection{Complexity Comparison}
\label{algorithm_complexity}

Quantum algorithms are characterized using three metrics: \textit{width, depth} and \textit{gate complexity}. Their values help us assess runtime, simplicity of implementation, and resilience to decoherence, which directly affect the effectiveness of the quantum algorithm. 
Moreover, the complexities of Regev's and Shor's algorithms are heavily influenced by the choice of multiplication algorithm. Depending on it, quantum metrics change significantly.

\begin{itemize}
\item \textit{Width} metric represents the number of qubits that are required to run an algorithm. It also takes into account ancillary qubits. The main purpose of ancillary qubits is to help perform computations -- they do not contain any input or output values. \textit{Width} affect the simplicity of implementation: the fewer the qubits, the easier it is to implement a quantum algorithm, due to the limited amount of qubits in existing quantum computers.

\item \textit{Depth} is the metric that describes the number of quantum gates in sequence. This value affects the quantum decoherence resistance as well as the run time of an algorithm. The fewer sequenced steps (gates), the faster and more robust the quantum circuit's performance. \textit{Depth} is often used to describe quantum algorithms interchangeably with \textit{width}, because it is common for one to be reduced at the expense of another.

\item \textit{Gate complexity} represents the quantity of operations (quantum gates) required to run the algorithm. Its value has an impact on run time as well as on the quantum decoherence resistance. The lower the \textit{gate complexity}, the less susceptible the algorithm's quantum effects are to destruction by external noise.
\end{itemize}



Shor's algorithm~\cite{shor1999} factorizes an $n$-bit integer using gate complexity of order $\widetilde{O}\left(n^2\right)$. But, as previously described, large gate complexity comes with a large decoherence and in the result the collapse of quantum effects. Regev's algorithm~\cite{regev2024} addresses this issue and proposes a solution with gate complexity of $\widetilde{O}\left(n^{3/2}\right)$. It should be pointed out that a quantum circuit of this size needs to be run $\sqrt{n}$ times. Moreover, the gate complexity can be reduced at the expense of the complexity of the classical part of the algorithm. This is the case if superpolynomial-time classical post-processing is allowed. In that case, for $1<\epsilon \leq 1/2$, the gate complexity is of order $\widetilde{O}\left(n^{3/2 - \epsilon}\right)$ with the classical post-processing part (solving the hard lattice problem) of order $\widetilde{O}\left(e^{n^{2\epsilon}}\right)$. Under these circumstances, the number of times that the quantum circuit needs to be run is $n^{1/2 + \epsilon}$. Regev states that the lower gate complexity comes at the expense of width, because his algorithm has a width of order $O\left(n^{3/2}\right)$, while the optimized Shor's algorithm $O(n)$. If it comes to depth, Shor's algorithm can be implemented with $\widetilde{O}\left(n^3\right)$~\cite{takahashi2006}, $\widetilde{O}(n)$~\cite{tan2024efficient} or even $\widetilde{O}(\log{n})$~\cite{cleve2000fast} depending on the optimization and width of the algorithm. Currently, an optimized Shor's algorithm uses depth of order $\widetilde{O}(n)$. Nevertheless, Regev states that his algorithm depth is smaller by $\widetilde{O}\left(n^{1/2}\right)$ than the original Shor's algorithm.

Currently, both Shor's and Regev's algorithms have been optimized. Table~\ref{tab:algorithms_metrics} presents a comparison between these two algorithms in terms of theoretical\footnote{The research paper~\cite{ragavan2024}, referenced in Table~\ref{tab:algorithms_metrics}, provides a theoretical and mathematical overview of the improvements and modifications to the implementation of Regev's algorithm. However, to our knowledge, there is no publicly available implementation of Regev's algorithm.} width and gate complexity based on the multiplication algorithm used. The comparison concerns just one run of the quantum algorithm: Shor's algorithm needs only $o(1)$ runs, while Regev's algorithm needs $o(\sqrt{n})$ runs in the proposed configuration.

In our work, we use $d \in \left\{\left\lceil \sqrt{n} \right\rceil, \left\lfloor \sqrt{n} \right\rfloor \right\}$ quantum input registers, each of them having width $\mathit{qd} \in \left\{\left\lceil \frac{n}{d} + d \right\rceil, \left\lfloor \frac{n}{d} + d \right\rfloor\right\}$. Our output register has width $n$, while an ancilla register uses $n + 1$ qubits. Therefore, the upper bound width of our algorithm is calculated as:

\[
O(d \cdot \mathit{qd} + n + n + 1) = O\left(\sqrt{n} \cdot 2\sqrt{n} + 2n + 1\right) = O(4n + 1) = O(n)
\]

As mentioned previously, Regev's algorithm is a multidimensional version of Shor's algorithm. The gate with the highest complexity metric upper bound value in both algorithms is a modulo exponentiation gate. In Shor's quantum circuit, there is only one modulo exponentiation gate, while in Regev's circuit, there are $d$ modulo exponentiation gates, but each of them uses a smaller number of qubits. Regardless of the algorithm, a modulo exponentiation gate consists of $2w$ multiplications and each multiplication consists of $w$ addition operations, where $w$ is the width of the input quantum register. For addition, we used H{\"a}ner implementation~\cite{stepien2021, haner2017}, which has gate complexity of $O(w\log{w})$. Here, $w$ denotes the widths of the quantum input and output registers. In Regev's algorithm, the input and output registers used by the modulo exponentiation gate have different widths, and because of this, we experienced difficulties in applying theoretical H{\"a}ner's complexity in our case. Thus, we decided to take into consideration the worst possible scenario and its complexity is of the order $O(w_{output} \log{w_{output}})$. 

As a result, if it comes to gate complexity in our implementation of Regev's algorithm, it consists of $d$ modulo exponentiation gates, each with $2\mathit{qd}$ multiplications, each with $\mathit{qd}$ operations of addition, each of complexity $O(n \log{n})$. Therefore, the upper bound gate complexity of our algorithm is calculated as follows:

\begin{equation*}
  \begin{matrix}
     O\left(d \cdot 2\mathit{qd} \cdot \mathit{qd} \cdot O(n\log{n})\right) = O(\sqrt{n} \cdot 2(2\sqrt{n}) \cdot 2\sqrt{n} \cdot O\left(n\log{n})\right) = \\
     = O\left(8n^{3/2} \cdot O(n \log{n})\right) = O\left(n^{3/2} \cdot O(n \log{n})\right) = O\left(n^{5/2}\log{n}\right)
  \end{matrix}
\end{equation*}

\begin{table}[h!]
\caption{Shor and Regev algorithms metrics comparison. The content is based on paper~\cite{ragavan2024}.}
\centering
\begin{tabular}{|c|c|c|c|}
\hline
\textbf{Algorithm} & \textbf{Multiplication algorithm} & \textbf{Width} & \textbf{Gate complexity} \\ \hline
Shor~\cite{shor1999} & Harvey, Hoeven~\cite{harvey2021} & $O\left(n\log{n}\right)$ &  $O\left(n^2\log{n}\right)$ \\ \hline
Shor & Schoolbook & $O(n)$ & $O\left(n^3\right)$ \\ \hline
Shor & Gidney~\cite{gidney2019asymptotically}, KMY~\cite{kahanamoku2024fast} & $O(n)$ & $O_\epsilon\left(n^{2+\epsilon}\right)$ \\ \hline
Optimized Shor~\cite{beauregard2003, takahashi2006, zalka2006, gidney2017factoring, haner2017} & Schoolbook &  $(1.5 + o(1))n$ & $\widetilde{O}\left(n^3\right)$ \\ \hline
Regev~\cite{regev2024} & Schoolbook & $O\left(n^{3/2}\right)$ & $\widetilde{O}\left(n^{3/2}\right)$ \\ \hline
Regev~\cite{regev2024} & Harvey, Hoeven~\cite{harvey2021} & $O\left(n^{3/2}\right)$ & $O\left(n^{3/2}\log{n}\right)$ \\ \hline
Optimized Regev~\cite{ragavan2024} & Harvey, Hoeven~\cite{harvey2021} & $O(n\log{n})$ & $O\left(n^{3/2}\log{n}\right)$ \\ \hline
Optimized Regev~\cite{ragavan2024} & Gidney~\cite{gidney2019asymptotically}, KMY~\cite{kahanamoku2024fast} & $(10.32 + o(1))n$ & $\widetilde{O}_\epsilon\left(n^{3/2+\epsilon}\right)$ \\ \hline
Optimized Regev~\cite{ragavan2024} & Schoolbook~\cite{roetteler2017} & $(10.32 + o(1))n$ & $O \left(n^{5/2}\log{n}\right)$ \\ \hline
Our implementation & Schoolbook & $O(n)$ & \makecell{$O \left(n^{5/2}\log{n}\right)$} \\ \hline

\end{tabular}

\label{tab:algorithms_metrics}
\end{table}

\FloatBarrier

The classical part of Shor's algorithm has computational complexity $O(n)$, where $n = \lceil \log_2{N} \rceil$. In case of Regev's algorithm, it is equal to the complexity of running Lenstra–Lenstra–Lovász (LLL) algorithm~\cite{lll} on the lattice $B$. 

By definition, for a given lattice $\mathcal{L}$ defined by the basis $F =\{f_1,...,f_z\}$, where $z \leq n$ and $f_1,...,f_z \in \mathbb{R}^n$, the LLL algorithm finds a set of nearly shortest vectors. In this case, the computational complexity of the LLL algorithm is equal to $O\left(z^5n\log^3 G\right)$, where $G = \max{(\lVert f_1 \rVert_2,...,\lVert f_z\rVert_2)}$. 

In our case, the lattice $B$ is a $d+m$ dimensional square lattice, that is, it has $d+m$ vectors in the base, which are members of set $\mathbb{R}^{d+m}$ and $M$ is the maximum Euclidean norm from the vectors in the lattice $B$. Therefore, the computational complexity of the classical part of Regev's algorithm is polynomial and equal to $O\left((d+m)^6\log^3 M\right)$. 


\section{Regev's Algorithm Implementation}
\label{cha:regev_algorithm_implementation}

Apart from the complexity of quantum and classical parts, the algorithm's performance is also affected by the hardware specifications, tools, chosen libraries, and used existing implementations.
For developing and running the quantum part of the algorithm, we used the open-source software development kit created by IBM Research, that is, Qiskit. We used three libraries provided by this SDK. The first was \texttt{qiskit} in version 1.2.4, which is used to create quantum circuits and to manage the classical-quantum data flow. The second library was \texttt{qiskit-aer} in version 0.15.1, which comes with a simulator that simulates quantum states on a classical computer, so in practice it is a quantum computer simulator. As for the third package library, it is a \texttt{quantum-ibm-runtime} library in version 0.32.0 and comes with the possibility of running the developed quantum circuit on IBM cloud quantum computers.

The authors of this paper built an implementation of Regev's algorithm based on the Stępień's implementation of Shor's algorithm available publicly in the Internet~\cite{stepien2021}\footnote{The implementation is available at~\url{https://github.com/bartek-bartlomiej/master-thesis} (accessed on 9 July 2025).}. In his work, the author implemented from scratch and compared the effectiveness of the Shor's algorithm using four different modular gates. These gates were created by Beauregard~\cite{beauregard2003}, Takahashi~\cite{takahashi2008}, H{\"a}ner~\cite{haner2017}. The fourth gate is Stępień's own combination of the gates mentioned above. Stępień implemented each of the more complicated quantum gates using primary quantum gates. Thus, we thought of his implementations as of the building blocks and used or modified some of them to create our own quantum circuit, as described in Section~\ref{sub:quantum_circuit}.

As for the classical part, in order to perform computation on lattices, we used a NumPy library. This library offers fast computation on matrices and vectors, thanks to the usage of NumPy arrays. They are faster and more compact than Python lists. Moreover, NumPy uses less memory to store data~\cite{NumPy}. Unfortunately, this library does not provide an implementation of the LLL algorithm. Therefore, we used an \texttt{olll} package, which is a dedicated Python tool providing an implementation of the LLL algorithm.

\subsection{Quantum Part}

The quantum part was developed using the Qiskit SDK. In addition, we used a quantum computer simulator provided with package \texttt{qiskit-aer}. It performs calculation on a \textit{statevector} that stores the probabilities of the basic states of qubits. Using this tool, the Qiskit SDK runs a quantum circuit once and collects many measurements. 

This feature is particularly beneficial in the context of Regev's algorithm. The quantum part theoretically requires $d+4$ executions. Using a quantum computer simulator, we simulate this behavior by performing a single execution of the quantum part, collecting $128$ measurements, and subsequently selecting $d+4$ vectors from the results rather than running a quantum part multiple times. The selection process of these $d+4$ vectors is performed using one of the three methods described above in Section~\ref{clas}. This approach significantly reduced the overall runtime of the algorithm, enabling a detailed investigation into the impact of various parameters of Regev's algorithm. In a real-world scenario, executing Regev's algorithm on a physical quantum computer could further optimize the runtime of the quantum component. Even if $d+4$ quantum part executions are required, this approach may still outperform the single-run simulation employed in this study.

Additionally, \texttt{qiskit-aer} package comes with different simulation methods, each with distinct properties. For example, some of them can support computation on a GPU or can simulate noise in quantum circuits. They also differ in the number of qubits they can simulate (excluding classical registers used for measurements) and in their hardware requirements. It is also possible to introduce noise to the simulator. In our implementation, we used the statevector simulator method, as it allows for simulating ideal circuit (without noise simulation) with the usage of CPU. However, this approach comes at the cost of large RAM consumption~\cite{qiskit2024}. A statevector of $n$-qubits uses $2^n$ complex values, each requiring 16 bytes of memory. This means that a 32-qubit vector uses 64\,GB of RAM. Our environment enabled us to perform computations on up to 29 qubits, which requires 8\,GB of RAM.

The applied quantum simulation method did not simulate quantum noise. In case of some of other simulation methods, or real quantum computers, it is necessary to apply quantum error mitigation techniques~\cite{cai2023}. Qiskit also provides error mitigation techniques for running a quantum circuit on real quantum computers with its \texttt{qiskit-ibm-runtime} package~\cite{error_mitigation_doc}.

\subsubsection{Parameters}
\label{quantum_part_parameters}
Regev's algorithm is a multidimensional version of Shor's algorithm. In our work, the number of dimensions depends on the value of the parameter $d$. Regev recommends its value to be $\left\lceil \sqrt{n} \right\rceil$ for the $n$-bit integer $N$. We decided to analyze two possible values for $d$ and those are $\left\lceil \sqrt{n} \right\rceil$ (later referenced as ``\texttt{ceil}'') and $\left\lfloor \sqrt{n} \right\rfloor$ (later referenced as ``\texttt{floor}''), so in the result $d \in \left\{\left\lceil \sqrt{n} \right\rceil, \left\lfloor \sqrt{n} \right\rfloor\right\}$. In practice, the discussed parameter denotes the number of quantum input registers and the number of squared primes. This in turn has a direct impact on the number of modulo exponentiation gates as well as the number of quantum Fourier transform gates and thereby on the quantum circuit width, depth and gate complexity. The complexity of the algorithm was discussed in Section~\ref{algorithm_complexity}.

In addition to the parameter $d$, we put in place the parameter $\mathit{qd}$. It can also take two possible values that result in $\mathit{qd} \in \left\{\left\lceil \frac{n}{d} + d \right\rceil, \left\lfloor \frac{n}{d} + d \right\rfloor\right\}$. Later, a version using the ceil function is referenced as ``\texttt{ceil}'' and the floor version as ``\texttt{floor}''. Thus, for example, the combination of the $d$ ceil version with the $\mathit{qd}$ floor version might be called \texttt{ceil\_floor}. Regev's recommendation for the parameter $\mathit{qd}$ is $\left\lceil \frac{n}{d} + d \right\rceil$ for an $n$-bit integer $N$. This parameter denotes the upper bound of the exponents $(z_1, z_2, ..., z_d)$ in the process of creating a superposition of a function. In case of a quantum circuit, the parameter $\mathit{qd}$ has an impact on the input register widths, and the widths of the quantum input registers affect the overall width, depth and gate complexity.

Due to the limitations of the quantum computer simulator and hardware specifications, there is a limited value of $N$ that we were able to factorize for \texttt{ceil\_ceil}, \texttt{ceil\_floor} parameters combinations. These limitations also affect Shor's algorithm, which runs for $N=77$ at the maximum for our simulations runs. Table~\ref{tab:parameters_max_N} illustrates the largest factorized number depending on the parameters $d$ and $\mathit{qd}$.

\begin{table}[h]
\caption{The biggest factorized number $N$ depending on $d$ and $\mathit{qd}$ parameters}
    \centering
    \begin{tabular}{|p{2cm}|p{2cm}||p{2cm}|}
        \hline
        \makecell{\centering $\textbf{d}$} & \makecell{\centering $\textbf{qd}$} & \makecell{\centering $\textbf{N}$} \\ \hline
        \makecell{\centering ceil} & \makecell{\centering ceil} & \makecell{\centering $57$} \\ \hline
        \makecell{\centering ceil} & \makecell{\centering floor} & \makecell{\centering $57$} \\ \hline
        \makecell{\centering floor} & \makecell{\centering ceil} & \makecell{\centering $119$} \\ \hline
        \makecell{\centering floor} & \makecell{\centering floor} & \makecell{\centering $143$} \\ \hline
    \end{tabular}
    \label{tab:parameters_max_N}
\end{table}

\subsubsection{Regev's Quantum Circuit}
\label{sub:quantum_circuit}

The implementation initializes a uniform superposition over the $d$ quantum input registers using Hadamard gates, which constitutes the first step in the quantum component of the algorithm. A Pauli-X gate is applied to the first qubit of the output register, which is equivalent to a NOT gate in classical computing. This operation sets the initial value of the output register to $1$, enabling it to store the result of the multiplication process in subsequent steps (by default, the output register holds a value of $0$). The next stage involves applying modulo exponentiation gates to each input register and the output register.

To implement this step, we utilized Stępień's adaptation of H{\"a}ner's modulo exponentiation gate. Regev's algorithm employs $d$ quantum input registers of smaller width $\mathit{qd}$, in contrast to Shor's algorithm, which uses a single quantum input register of width $n$. Consequently, we modified H{\"a}ner's modulo exponentiation gate by eliminating redundant quantum input qubits in each of the $d$ gates.
In addition, we adopted Stępień's naming convention~\cite{stepien2021} for quantum gates. The modulo exponentiation gate is referred to as the ``Exp'' gate and is decomposed into $\mathit{qd}$ multiplication gates, denoted as ``C-U'' gates. Following this, the output register is ignored (measured), as its value is not relevant to our objectives. The next step involves applying Quantum Fourier Transform (QFT) gates to each quantum input register. Finally, the values of the quantum input registers are measured and stored in the classical registers.



\begin{figure}
    \centering
    \includegraphics[width=0.8\linewidth]{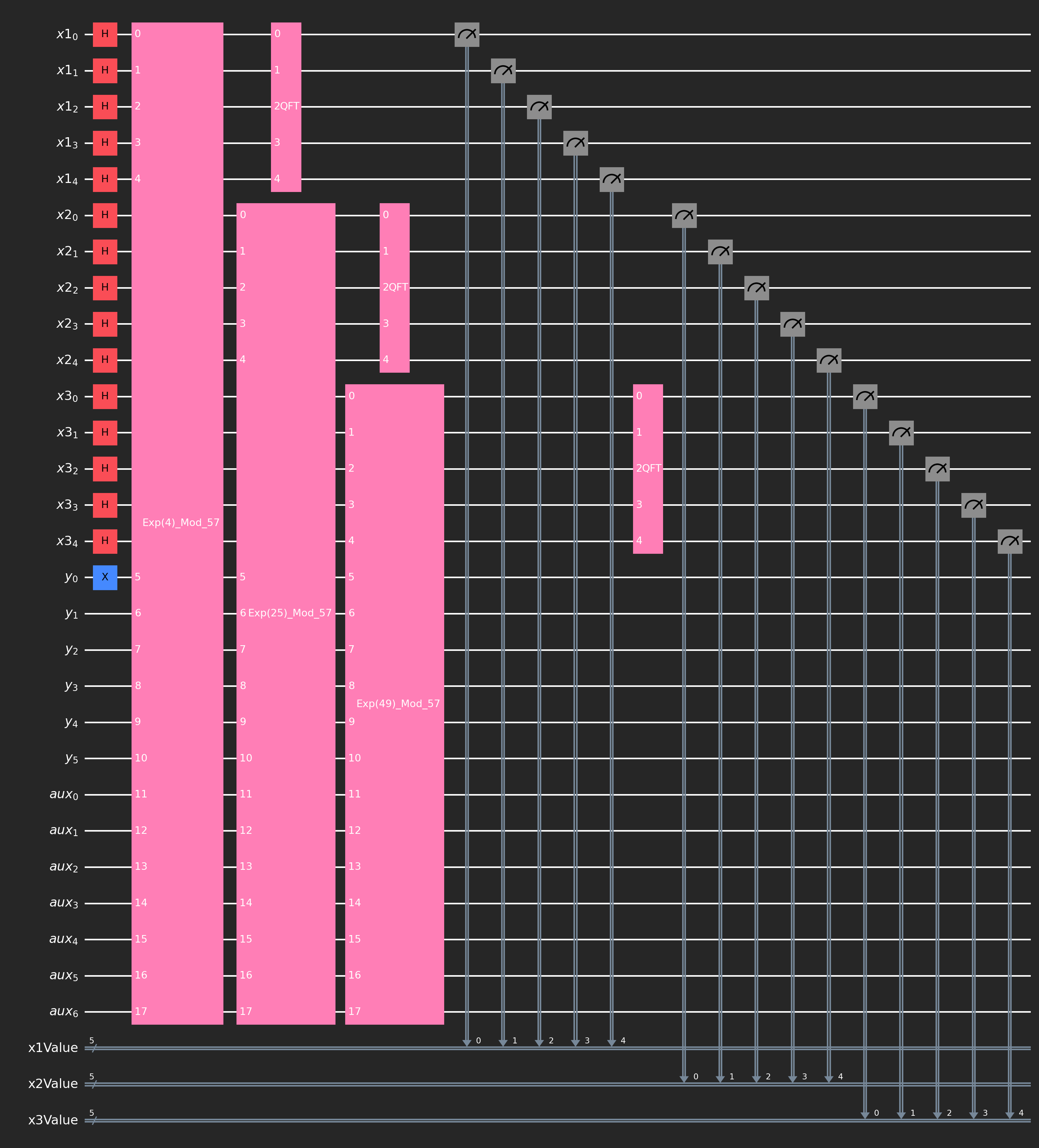}
    \caption{General quantum circuit for $N=57$, $d$ \texttt{ceil} and $\mathit{qd}$ \texttt{ceil} (\texttt{ceil\_ceil}) parameters}
    \label{fig:57_ceil_ceil_general}
\end{figure}

\begin{figure}
    \centering
    \includegraphics[width=1\linewidth]{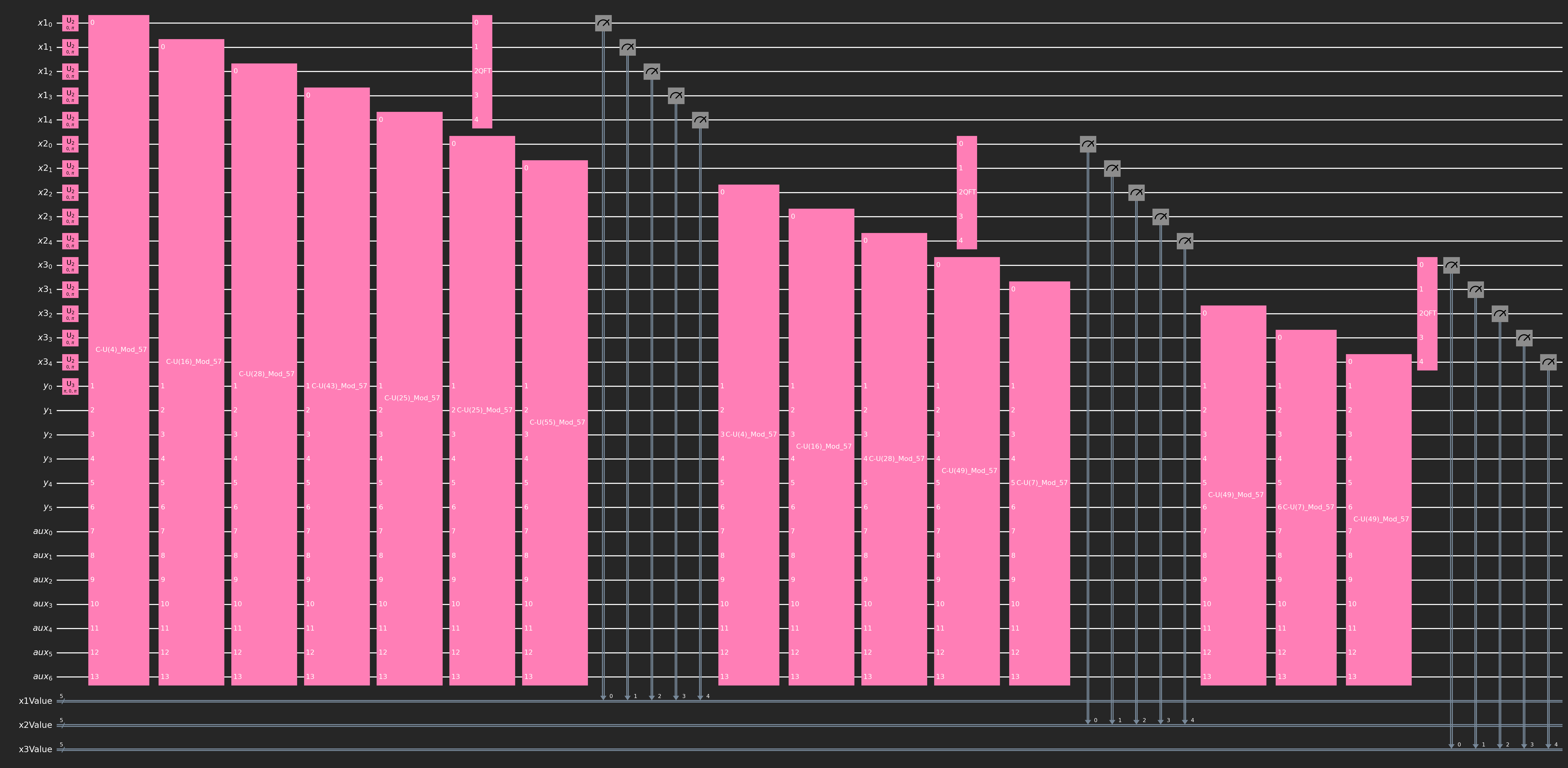}
    \caption{Decomposed quantum circuit for $N=57$, $d$ \texttt{ceil} and $\mathit{qd}$ \texttt{ceil} (\texttt{ceil\_ceil}) parameters}
    \label{fig:57_ceil_ceil_decomposed}
\end{figure}

We noted that the ``Exp'' modular exponentiation gate decomposes into $\mathit{qd}$ modular multiplication gates, referred to as ``C-U'' gates. To illustrate the impact of this decomposition, we present the quantum circuits in two forms: general and decomposed. This approach emphasizes that any modification leading to the addition of a single ``Exp'' quantum gate in the circuit actually results in an increase in the number of underlying quantum gates that make up that additional gate.
Figures~\ref{fig:57_ceil_ceil_general} and~\ref{fig:57_ceil_ceil_decomposed} depict a general and decomposed quantum circuits accordingly. The factorized number is $57 = 3 \cdot 19$ and a parameter combination is \texttt{ceil\_ceil} (parameters $d$ and $\mathit{qd}$ are taken in the \texttt{ceil} version, which results in $d=3$ and $\mathit{qd}=5$). It is worth mentioning that in reality, the ``Exp'' gates are small quantum circuits. This also holds for the ``C-U'' gates, which are composed of the modulo adder gates. These, in turn, are composed of gates such as adder or carry, and these gates are finally composed from primary gates such as Pauli-X (X) or Hadamard gates. These gates compositions are described in detail in Stępień's work~\cite{stepien2021}.

\subsubsection{Output Vectors}
\label{sub:output_vectors}

Applying the Quantum Fourier Transform (QFT) and measuring the states of each quantum input register yields a specific output vector. For example, the output vector $[10, 22, 21]$ consists of three values, each corresponding to the measurement of a quantum input register at the end of the quantum circuit. Measurement of the first quantum input register produces the value \(10\), the second register yields \(22\), and the third register results in \(21\).

In Qiskit's nomenclature, measurements are referred to as \textit{shots}. Due to the probabilistic nature of qubits in a quantum superposition, the values in the output vector may vary across multiple runs of the quantum circuit. 
Another consequence of qubit probabilistic behavior is that certain vectors have a higher probability of occurrence than others. As a result, repeated measurements of the quantum circuit output reveal variations in the frequency of observed vectors. This phenomenon can be seen, for example, in Figure~\ref{fig:vect_N_51_ceil_ceil}, where the vector $[20, 12, 8]$ appears the least frequently (has only $9$ shots), while the vectors $[8, 24, 16]$, $[12, 20, 24]$, and $[28, 4, 24]$ occur most frequently (these vectors have $20$, $21$ and $21$ shots, respectively), indicating their higher probability of being measured.

In summary, the set of measured vectors depends on the factorized number $N$ and the combination of parameters. The frequency with which a measurement produces the same vector (vector occurrences) is determined by its probability. Consequently, even for the same $N$ and combination of the parameters, both the set of output vectors and their occurrences may vary between different measurement collections.

In our implementation, we collected 128 measurements for each $N$ using a single quantum circuit run. Consequently, the total number of vector occurrences sums up to 128. For example, the difference between Figures~\ref{fig:vect_N_51_ceil_ceil} and~\ref{fig:vect_N_51_ceil_floor} illustrates the impact of quantum probabilities on the occurrences of the output vectors for $N = 51$ with the \texttt{ceil\_ceil} and \texttt{ceil\_floor} parameter settings.

\begin{figure}[h!]
\centering
\begin{BVerbatim}
vector with 9 shots: [20, 12, 8]
vector with 11 shots: [24, 8, 16]
vector with 13 shots: [4, 28, 8]
vector with 16 shots: [16, 16, 0]
vector with 17 shots: [0, 0, 0]
vector with 20 shots: [8, 24, 16]
vector with 21 shots: [12, 20, 24]
vector with 21 shots: [28, 4, 24]
\end{BVerbatim}
\caption{Output vector measurements for $N=51$ and \texttt{ceil\_ceil} parameters.}
\label{fig:vect_N_51_ceil_ceil}
\end{figure}

\begin{figure}[h!]
    \centering
    \begin{BVerbatim}
vector with 13 shots: [28, 4, 24]
vector with 14 shots: [16, 16, 0]
vector with 15 shots: [12, 20, 24]
vector with 15 shots: [0, 0, 0]
vector with 16 shots: [4, 28, 8]
vector with 17 shots: [8, 24, 16]
vector with 19 shots: [24, 8, 16]
vector with 19 shots: [20, 12, 8]
    \end{BVerbatim}
    \caption{Output vector measurements for $N=51$ and \texttt{ceil\_floor} parameters.}
    \label{fig:vect_N_51_ceil_floor}
\end{figure}

\subsection{Classical Part}
\label{clas}

The goal of the classical part of Regev's algorithm is to process the output of a quantum circuit to factorize a given number $N$. 
The environment for computations for the classical part was the same as for the quantum part. We used CPU with 12 logical cores (6 cores with 2 logical threads per core) and RAM with the total size of 16\,GB. In our implementation, we used the parameters proposed in Kiebert's work~\cite{midas2024}. However, we rounded the parameter $S$ to the nearest integer. Without this adjustment, the LLL algorithm for some numbers could reduce the vectors to a zero vector. The parameter $T$ is defined as the upper bound of the norm of the smallest vector in $\mathcal{L} \setminus \mathcal{L}_0$. For small numbers, it is possible to find exactly such a vector with complexity $O\left(\mathit{\mathit{qd}}^d\right)$, but it is equivalent to finding a result vector that allows computing $p$ and $q$ for $N$ without a usage lattice and LLL algorithm. For large numbers, this approach is very time consuming. To make our solution scalable, we approximate the value of $T$ using a heuristic assumption introduced in Regev's work. It states that there exists a vector in ${\mathcal{L} \setminus \mathcal{L}_0}$ with the norm at most $T = \exp{\left(O(n/d)\right)}$. This assumption provides an estimate for the likely norm of the desired vectors before applying the LLL algorithm. Based on this, we can predict the value of $R$, which influences the determinant of the matrix $B$. By estimating $R$, we can adjust the determinant of $B$ to ensure that, after applying the LLL algorithm, we are likely to obtain vectors with the desired norm\footnote{The correlation between the determinant of matrix $B$ and the impact on the length of vectors after applying the LLL algorithm, along with a mathematical proof, is detailed in Kierbert's work~\cite{lll}.}. For small $N$ ($N \leq 57$), we found that a good approximation for $T$ is $\left\lceil \exp{\left(\frac{n}{2d}\right)} \right\rceil$.  This result was obtained by calculating accurate values of $T$ for a few cases and determining the best-fitting curve. The comparison between the exact values of $T$ and the approximation using the function $\exp{\left(\frac{n}{2d}\right)}$ is presented in Table~\ref{tab:N_to_T}.

 \begin{table}[h]
     \caption{Comparison of exact values of $T$ and their approximations using function $\exp{\left(\frac{n}{2d}\right)}$ for selected values of $N$}
    \centering
    \begin{tabular}{|c|c|c|c|c|c|}
        \hline
        $N$                & 15 & 21 & 35 & 51 & 57 \\ \hline
        value of $n$       & 4  &  5 &  6 & 6 & 6 \\ \hline
        value of $d$       & 2  & 3  & 3 &  3 & 3 \\ \hline
        exact value of $T$ & 3  & 2  & 3 & 3 & 3 \\ \hline
        approximate value of $T$ &  2.718 & 2.301 & 2.718 &2.718 &2.718 \\ \hline
    \end{tabular}
    \label{tab:N_to_T}
\end{table}

As described earlier, the time-optimal method for collecting multiple output vectors from the quantum circuit consists in executing the circuit once and then performing multiple measurements of the \textit{statevector}, rather than running the quantum circuit independently multiple times. Due to the features of the Qiskit library, this approach does not have an impact on the accuracy of the results. Therefore, we collect 128 vectors for every parameter (\texttt{ceil\_ceil}, \texttt{ceil\_floor}, \texttt{floor\_ceil} and \texttt{floor\_floor}) and for all values $N$ that our hardware allows us to simulate (as indicated in Table~\ref{tab:parameters_max_N}). Therefore, after measuring quantum output vectors, we simulate many independent runs of circuits. In our program, we implemented a method called \texttt{run\_file\_data\_analyzer}, which analyzes the outputs of quantum circuit saved in files. The purpose of this function is to measure the effectiveness of finding square roots of unity modulo-$N$ (highlighting non-trivial ones) for our implementation of Regev's algorithm. This method takes four parameters. Two of them specify which quantum output files to analyze, while the other two define the type of test (parameter \texttt{type\_of\_test}) and the number of repetitions (parameter \texttt{number\_of\_combination}). We implemented three different types of tests.

In the first test, we randomly took $d+4$ vectors according to the probability of their return by the quantum circuit. This approach allows us to simulate $d+4$ executions of the quantum circuit.

In the second test, we wanted to check whether there exist some vectors that, despite frequent occurrence among quantum circuit output vectors, might not be a good approximation of the vectors from the dual lattice $\mathcal{L}^*$. To achieve that, we selected each vector from the quantum circuit's output with equal probability to form a set of $d+4$ vectors, regardless of how many times a specific vector was returned. If the results of the second test were better than those of the first test, it would indicate that certain frequently returned vectors -- such as $[0,0,0]$ for $N=21$ -- reduce the algorithm's effectiveness.

In the third test, we generated completely random vectors. The length of the vector is defined by parameter $d$. In the quantum circuit, each coordinate of the vector is measured from $\mathit{qd}$ qubits. That means that if it was unknown what happens in the circuit, we could assume that the measurement would return values from the range $[0, 2^{\mathit{qd}})$. Thus, according to this method, we create vectors of length $d$ with random coordinates in the range $[0, 2^{\mathit{qd}})$. This test allows us to verify whether the correct factoring of $N$ in our implementation of Regev's algorithm is not a coincidence. 

In each test, we measured two parameters. The first parameter is a percentage of vectors that, after computation, returns the square root of unity modulo-$N$. The second parameter is the percentage of vectors that return non-trivial square roots of unity modulo-$N$, i.e., values not equal to $-1$ or $1$. For every number $N$ and parameters $d$ and $\mathit{qd}$ that we tried to factorize on the quantum circuit simulator, we ran the three tests 1000 times to gather sufficient statistics.


\section{Comparative Analysis}\label{an1}

Following an implementation of the algorithm, we conducted a series of experiments. The research focused on the runtime and efficiency depending on the various combinations of parameters. After this, we collected data about the Shor's algorithm and compared it with the Regev's performance and efficiency. 
In the illustrated point-line plots, the points represent the actual factorized numbers and their corresponding runtime or efficiency. Lines were added to facilitate comparison between algorithm runs for different parameter values. The set of numbers below presents all the factorized numbers $N$:

\begin{equation}
15, 21, 33, 35, 39, 51, 55, 57, 65, 69, 77, 85, 91, 95, 119, 143
\label{eq:factorized_N}
\end{equation}

\subsection{Runtime Analysis}

This section presents the Regev's algorithm quantum and classical part run-time performance for all analyzed parameters configurations and their comparison with the Shor's algorithms runtime. We used a Qiskit SDK with a quantum computer simulator and also found that it is sufficient to run a quantum circuit once. 
To collect run-time data, the quantum and classical parts of the algorithms were executed only once. This approach was chosen to focus on the impact of parameter variations and due to the time constraints.
In practice, the run-time of a quantum part on a real quantum computer should be considerably faster.

In Section~\ref{quantum_part_parameters}, we mentioned that for each of the parameters combination there was a maximum factorized number $N$ (Table~\ref{tab:parameters_max_N}). This is the reason why the data series on the following graphs have different maximum values of $N$ on the X-axis. Also, we decided to analyze the quantum and classical parts separately because the runtime of the classical parts in Regev's and Shor's algorithms is negligibly small (the classical part runs in about half a second, while the quantum part runs for minutes to hours). Thus, the comparison of the algorithms' speeds corresponds to the comparison between their quantum parts.

\begin{figure}[h!]
\centering
\includegraphics[width=0.8\textwidth]{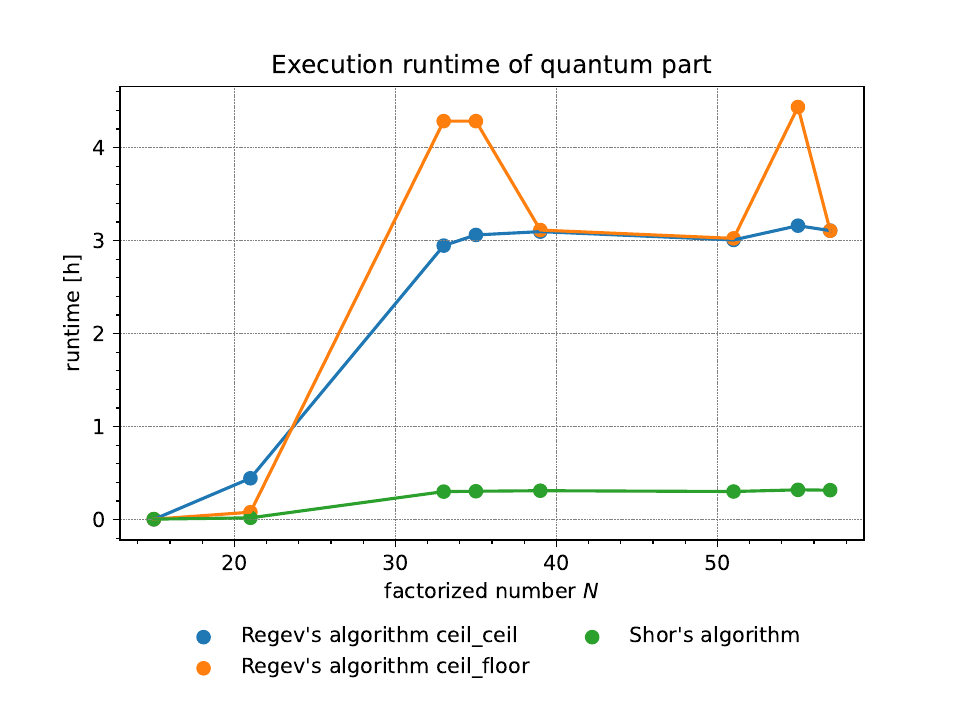}
\caption{Runtime comparison for \texttt{ceil\_ceil} and \texttt{ceil\_floor} parameters}
\label{fig:quantum_chart_ceils_and_shor}
\end{figure}

Figure~\ref{fig:quantum_chart_ceils_and_shor} presents the runtime of the Regev's algorithm for $d$ parameter in \texttt{ceil} version along with the performance of Shor's algorithm. For an analyzed $N$ range, Shor's algorithm is considerably faster, which aligns with Regev's remark that his algorithm is asymptotically faster than Shor's algorithm. This means that there exists a value of $N$ for which Regev's algorithm outperforms Shor's algorithm in terms of speed, although this value might be very large. It can be seen that for the number $N=33$ runtime for all presented runs increased significantly. This is because the previous number $N=21$ is an integer of $n=5$ bits, while $N=33$ is an integer of $n=6$. The number that is $1$ bit larger affected quantum output and ancillary registers by adding one qubit, so it increased the width of a quantum circuit by two qubits and resulted in an increase in the runtime of the quantum part of the algorithm.

\begin{figure}
\centering
\includegraphics[width=0.8\textwidth]{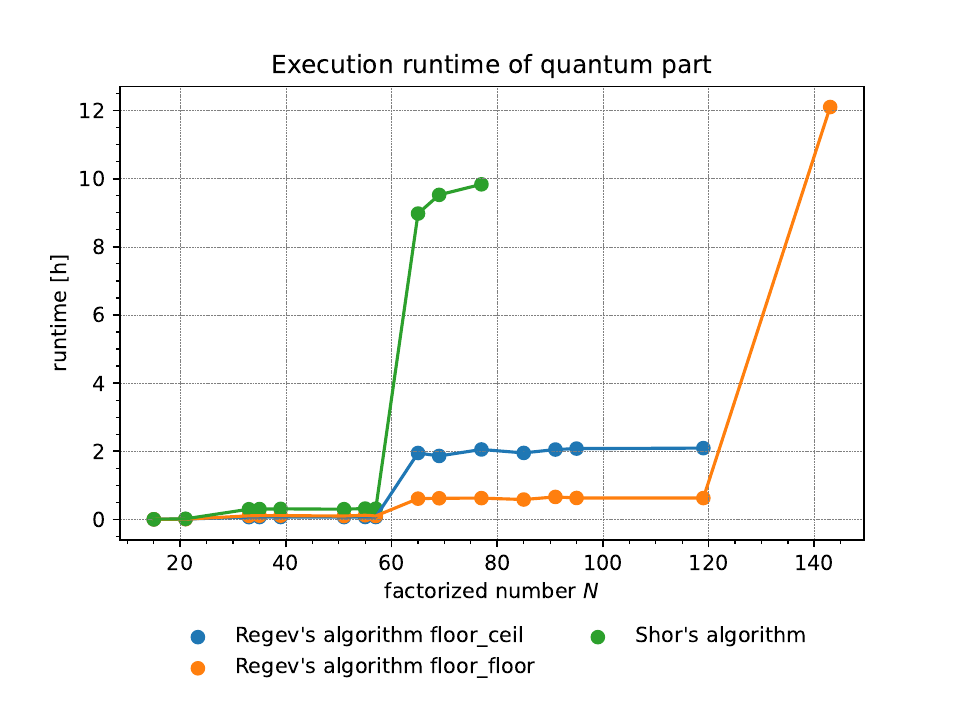}
\caption{Quantum part runtime comparison for \texttt{floor\_ceil} and \texttt{floor\_floor} parameters}
\label{fig:quantum_chart_floors_and_shor}
\end{figure}

Figure~\ref{fig:quantum_chart_floors_and_shor} depicts a comparison between Regev's algorithm with $d$ parameter in \texttt{floor} version and Shor's algorithm. Contrary to the previous graph, here the maximum $N$ is equal to $143$ for \texttt{floor\_floor} parameters. In the case of Shor's algorithm, the maximum factorized number was $77$. In the same way as before, there is a significant increase in run-time for number $N=65$ regarding all of the runs. Similarly, $N=65$ is $1$ bit larger than previously analyzed number $N=57$. An analogous situation can be observed in the case of the number $N=143$ for \texttt{floor\_floor} parameters. For this set of parameters, Shor's algorithm is slower, but this is at the cost of the efficiency of the Regev's algorithm.

For the ranges and parameters of the analyzed number $N$ and the parameters, in terms of speed, the best performance has parameters $d$ in the \texttt{floor} and $\mathit{qd}$ in \texttt{floor} modes. This set of parameters indicates the smallest number of dimensions as well as the smallest exponentiation boundary among the analyzed parameters. In the result, this set of parameters directly affects a number of quantum input registers, as well as the width of the quantum circuit. The conclusion is that under the given conditions regarding $N$ and the parameters, fewer dimensions and smaller exponentiation boundaries result in a faster quantum part runtime of Regev's algorithm. Moreover, for this set of parameters, Regev's algorithm was able to factorize a bigger maximum number $N$ than the other parameters.

\begin{figure}
\centering
\includegraphics[width=0.8\textwidth]{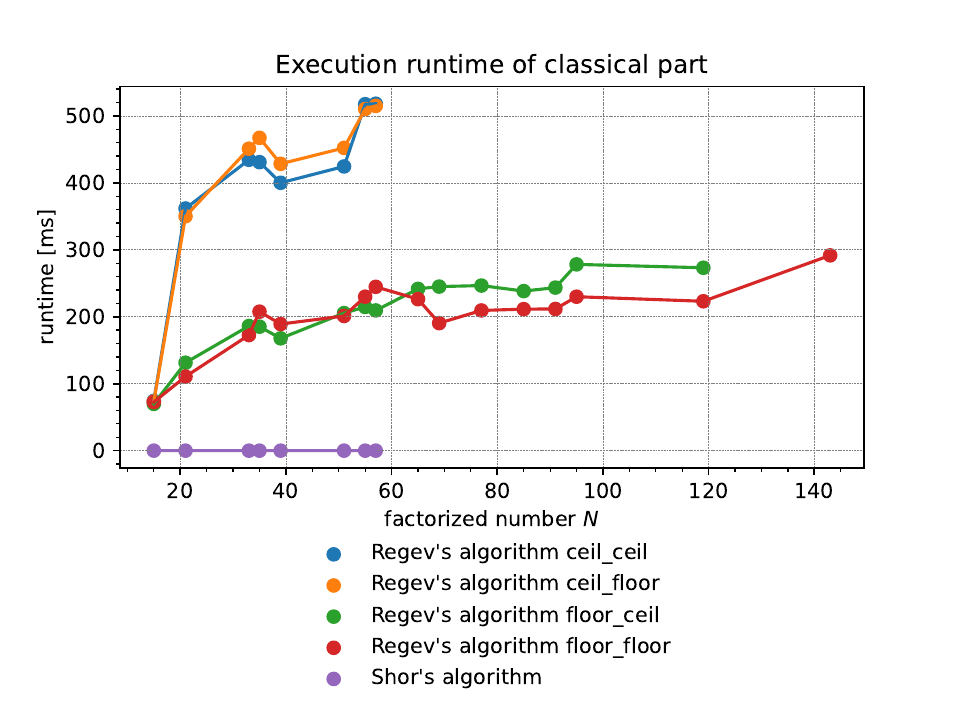}
\caption{Classical part runtime comparison for all parameters configurations}
\label{fig:classical_chart_all_and_shor}
\end{figure}

The classical computations underlying Regev’s algorithm exhibit greater complexity than those of Shor’s algorithm. Consequently, we compared the runtimes of these classical components to evaluate how these differences might affect the overall computation time.
Figure~\ref{fig:classical_chart_all_and_shor} illustrates the run-time of classical computations for Regev's and Shor's algorithms. It can be seen that Shor's algorithm's classical part is significantly faster than Regev's classical part, but considering a quantum part runtime, it is negligible for the whole algorithm's speed. When comparing Regev's algorithm parameters, the faster are those with $d$ parameter in the \texttt{floor} mode. It can be observed that the $\mathit{qd}$ parameter does not have a significant impact on the algorithm's classical parts' speed. Similarly to the execution time of quantum parts, the best performance among the combinations of parameters analyzed is for $d$ in the \texttt{floor} and $\mathit{qd}$ in \texttt{floor} modes.

\subsection{Efficiency Analysis}
\label{efficiency}

In this section, we present the results of the effectiveness of our implementation of Regev's algorithm. To automate the testing process, we implemented the function \texttt{run\_file\_data\_analyzer}, as described in Section~\ref{clas}. We define one test run as taking $d+4$ vectors using the method specified by the parameter \texttt{type\_of\_test}, constructing a matrix $B$ from these vectors, executing the LLL algorithm, and verifying two properties. The first property checks whether there exists a vector among the returned ones that allows recovering a square root of unity modulo-$N$, i.e., whether the returned vector belongs to the lattice $\mathcal{L}$. The second property checks whether there exists a vector among the returned ones that allows recovering a non-trivial square root of unity modulo-$N$, i.e., whether the returned vector belongs to the lattice $\mathcal{L} \setminus \mathcal{L}_0$.

Firstly, we tested the effectiveness of our implementation using parameters defined in Kierbert's work~\cite{midas2024} to create matrix $B$.
We measured the effectiveness for all values of $N$, where we ran the quantum circuit, using all combinations of rounding for $d$ and $\mathit{qd}$, i.e., \texttt{ceil\_ceil}, \texttt{ceil\_floor}, \texttt{floor\_ceil}, and \texttt{floor\_floor}. For these parameters, we performed the three types of tests, which were thoroughly described in Section~\ref{clas}. For each rounding method, we prepared the graph which presents the results for all three combined tests, allowing for easier comparison of the test outcomes.

\begin{figure}
  \centering
  \includegraphics[width=0.8\textwidth]{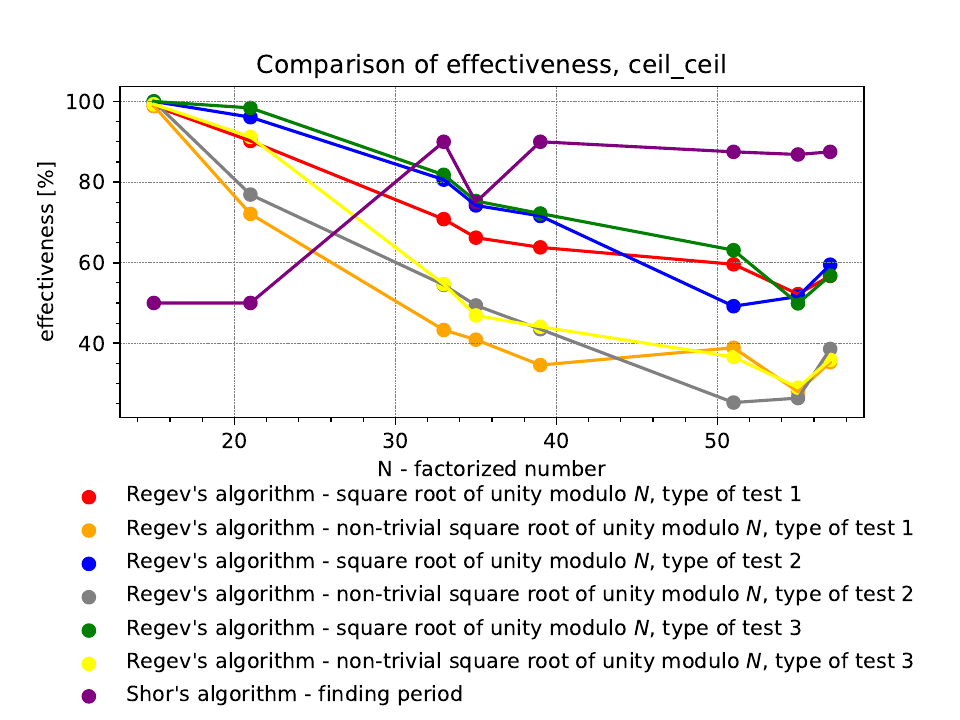}
  \caption{Effectiveness comparison for \texttt{ceil\_ceil} for all types of test}
  \label{fig:effectiveness_ceil_ceil_all}
\end{figure}

The results of the tests for the parameter \texttt{ceil\_ceil} are shown in Figure~\ref{fig:effectiveness_ceil_ceil_all}. It is evident that for the first type of test, the efficiency of finding square roots of unity modulo-$N$ (both trivial and non-trivial) declines as the factorized number $N$ increases. For the second type of test, the effectiveness decreases up to $N=51$, after which it increases slightly for $N=55$ and $N=57$. In the third type of test, the effectiveness decreases up to $N=55$ and then increases for $N=57$. For the parameter \texttt{ceil\_ceil}, the highest efficiency was achieved in the third type of test, although the efficiency for the second test was nearly identical, except for $N=51$. The first test exhibited the lowest efficiency. The general shape of the plots across all tests for parameter \texttt{ceil\_ceil} is similar.

\begin{figure}
  \centering
  \includegraphics[width=0.8\textwidth]{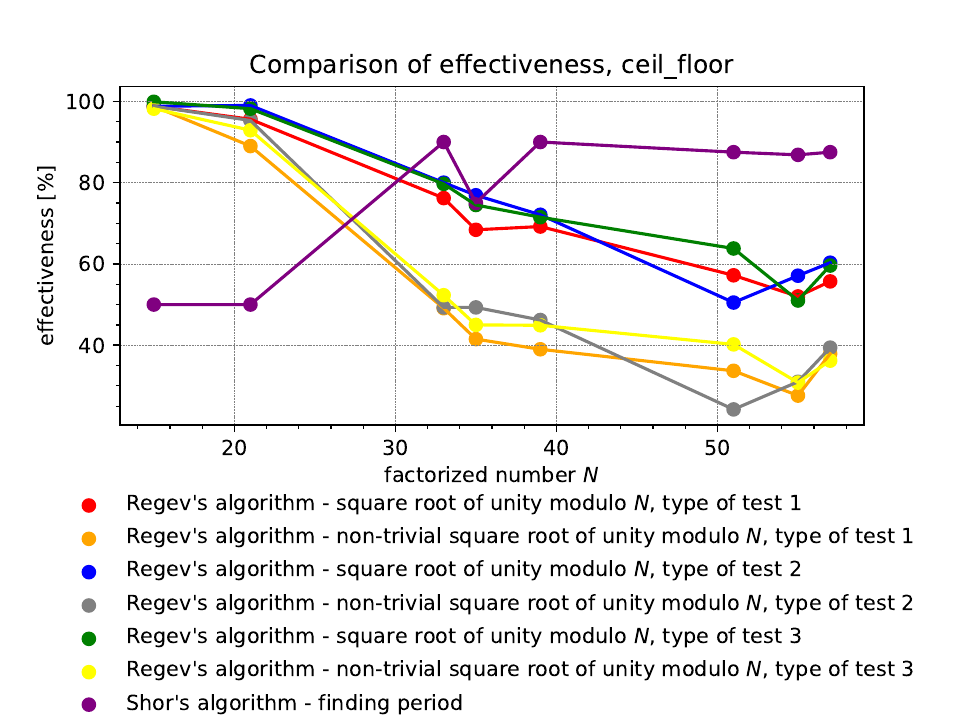}
  \caption{Effectiveness comparison for \texttt{ceil\_floor} for all types of test}
  \label{fig:effectiveness_ceil_floor_all}
\end{figure}

The results of the tests for the parameter \texttt{ceil\_floor} are shown in Figure~\ref{fig:effectiveness_ceil_floor_all}. The results are very similar to those obtained for the parameter \texttt{ceil\_ceil}, with the shapes of the figures being almost identical. We observe increases and decreases for the same values of the parameter $N$. However, the effectiveness of all three tests shows more consistent values compared to the results for the parameter \texttt{ceil\_ceil}.

For both parameters, \texttt{ceil\_ceil} and \texttt{ceil\_floor}, we observe that for $N=15$ and $N=21$, the efficiency of factorizing $N$ using our implementation of Regev's algorithm is better than that of Shor's algorithm. However, after $N=33$, Shor's algorithm stabilizes at an efficiency of approximately 88\%, while Regev's algorithm exhibits a declining trend and performs significantly less efficiently than Shor's algorithm.

\begin{figure}
  \centering
  \includegraphics[width=0.8\textwidth]{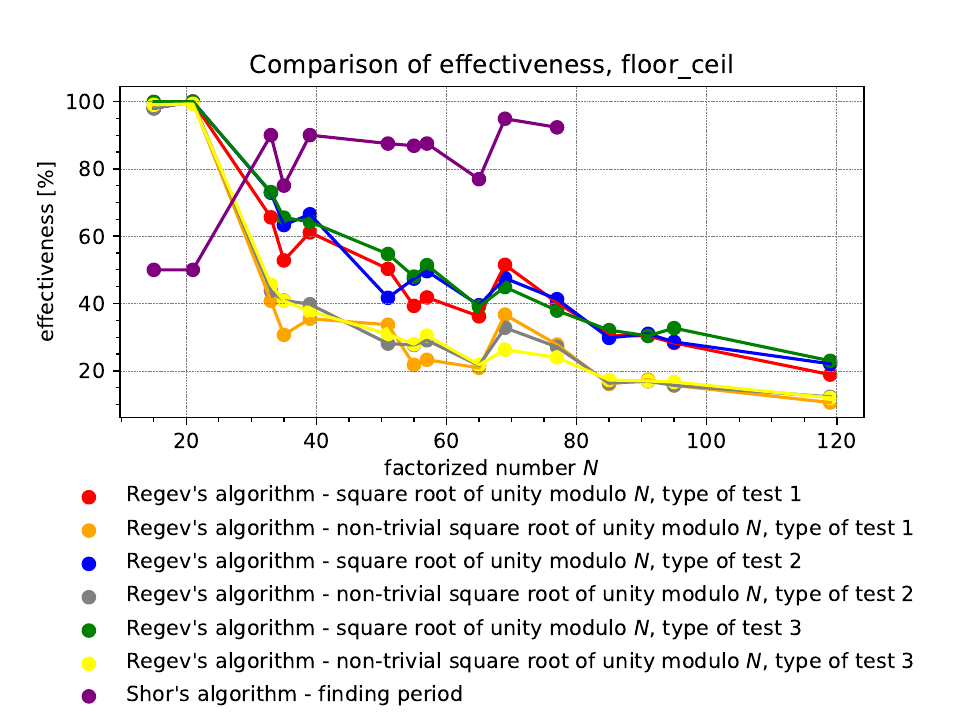}
  \caption{Effectiveness comparison for \texttt{floor\_ceil} for all types of test}
  \label{fig:effectiveness_floor_ceil_all}
\end{figure}

The results of the tests for the parameter \texttt{floor\_ceil} are shown in Figure~\ref{fig:effectiveness_floor_ceil_all}. All three tests have a similar shape in their figures. For the first type of test, we observe a decreasing trend, with a single peak at $N=51$. For the second and third types of test, a similar declining trend is evident, but the results show smaller peaks, with the third test having the smallest peaks.

\begin{figure}
  \centering
  \includegraphics[width=0.8\textwidth]{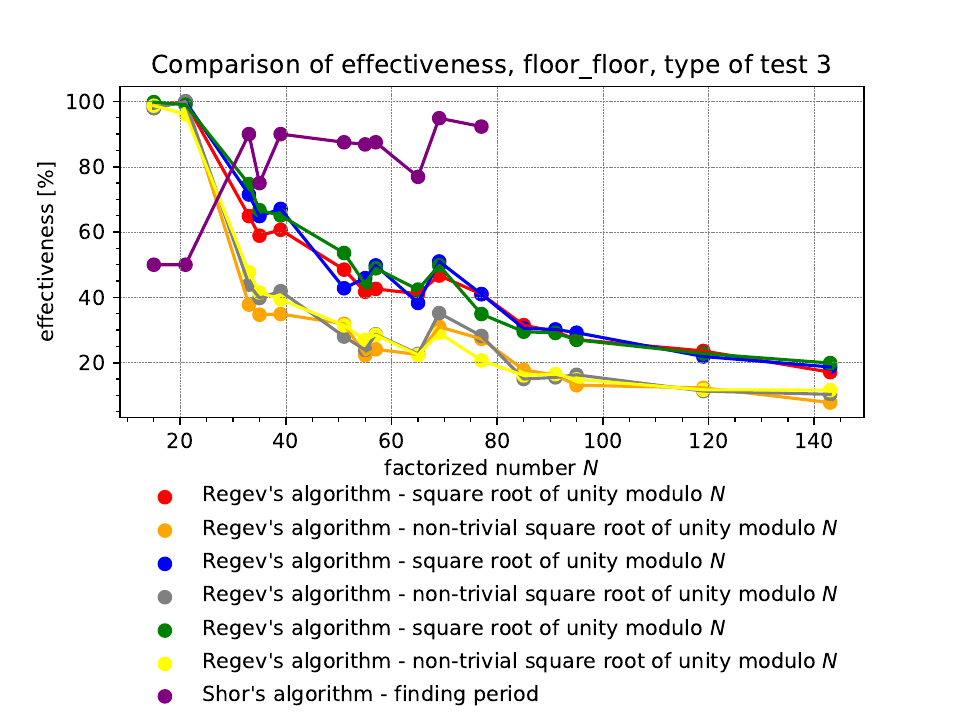}
  \caption{Effectiveness comparison for \texttt{floor\_floor} for all types of test}
  \label{fig:effectiveness_floor_floor_all}
\end{figure}

The results of the tests for the parameter \texttt{floor\_floor} are shown in Figure~\ref{fig:effectiveness_floor_floor_all}. For the first type of test, we observe a declining trend, with a single peak at $N=69$. For the second and third types of tests, the shapes of the functions are very similar, with two significant peaks in effectiveness at $N=57$ and $N=69$. Both the second and third tests have a decreasing trend.

For the parameters \texttt{floor\_ceil} and \texttt{floor\_floor}, we obtained worse effectiveness compared to \texttt{ceil\_ceil} and \texttt{ceil\_floor}. However, the choice of the rounding parameter $d$ has a greater impact on effectiveness than the rounding of the parameter $\mathit{qd}$. The rounding of $\mathit{qd}$ does not appear to significantly influence efficiency. Different rounding schemes for $\mathit{qd}$, when combined with the same rounding for $d$, result in situations where \texttt{ceil} performs better in some cases, while \texttt{floor} performs better in others. The declining trend in the efficiency of our implementation of Regev's algorithm as $N$ increases is not observed in Shor's algorithm, which stabilizes and maintains high efficiency. This decreasing tendency in our implementation of Regev's algorithm is consistent across all three tests and for all parameters (\texttt{ceil\_ceil}, \texttt{ceil\_floor}, \texttt{floor\_ceil} and \texttt{floor\_floor}). This consistency suggests that the parameters chosen in the classical part may not be optimal. Furthermore, we do not know the reason for the observed peaks for some values of $N$ and declines for others.

The above results indicate that our implementation, based on parameters defined in Kierbert's work, is not efficient. Therefore, we attempted to find better parameters for the classical part. We focus our tests on the parameter $t$, which is used in the quantum part to create the Gaussian distribution. However, in our implementation, we utilized a uniform superposition instead of Gaussian distribution. Thus, the parameter $t$ was used exclusively in the classical part to transform the output vectors from the quantum circuit into approximations of vectors in the lattice $\mathcal{L}^*$.

We hypothesized that, since the parameter $t$ is not used in the quantum part to create a Gaussian distribution, adjusting its value in the classical part might yield better results. We concentrated our research on the parameter \texttt{ceil\_ceil}, as this value was proposed in Regev's work. We conducted the first type of test for this parameter, exploring $t$ values in the range $[2, 20]$, and attempted to identify the optimal value of $t$. 

\begin{figure}[h!]
  \centering
  \includegraphics[width=0.8\textwidth]{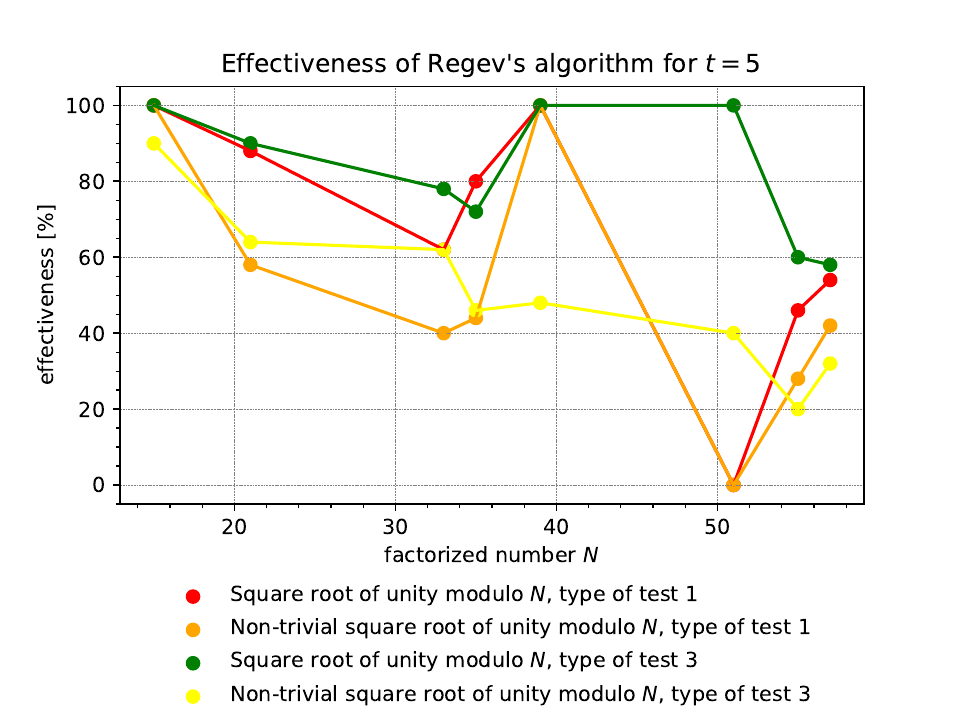}
  \caption{Effectiveness comparison for $t=5$}
  \label{fig:effectiveness_t_5}
\end{figure}

\begin{figure}[h!]
  \centering
  \includegraphics[width=0.8\textwidth]{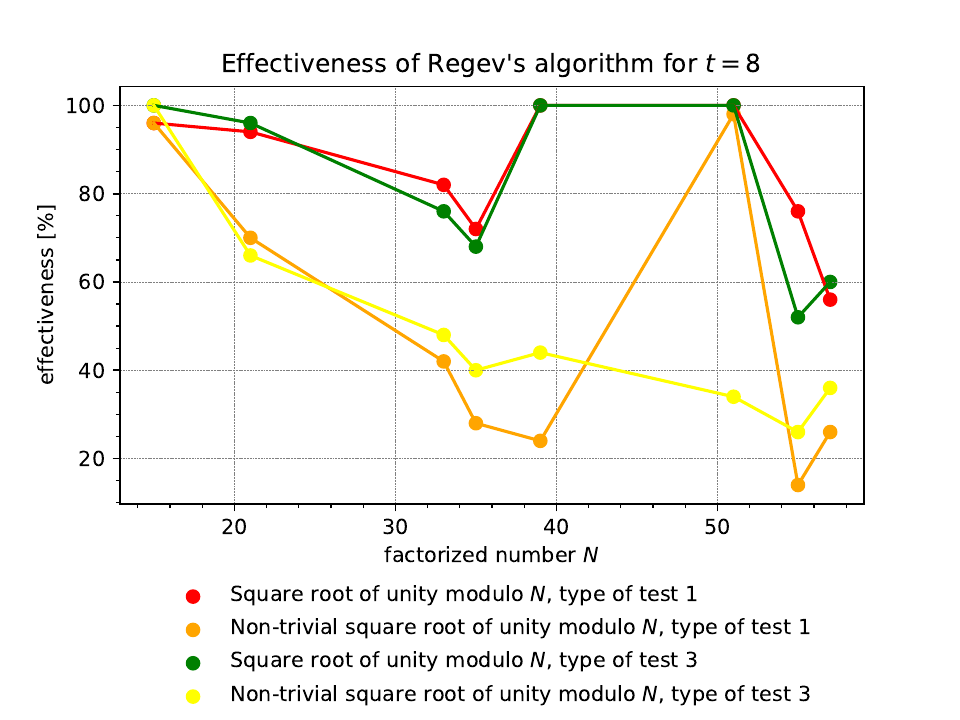}
  \caption{Effectiveness comparison for $t=8$}
  \label{fig:effectiveness_t_8}
\end{figure}

\begin{figure}[h!]
  \centering
  \includegraphics[width=0.8\textwidth]{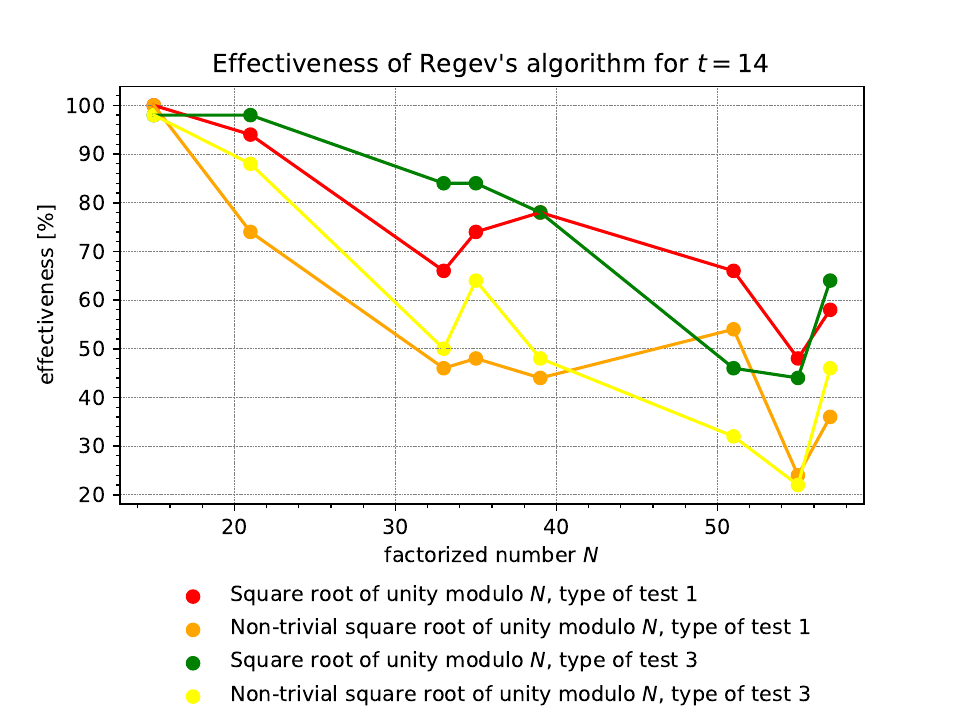}
  \caption{Effectiveness comparison for $t=14$}
  \label{fig:effectiveness_t_14}
\end{figure}

\begin{figure}[h!]
  \centering
  \includegraphics[width=0.8\textwidth]{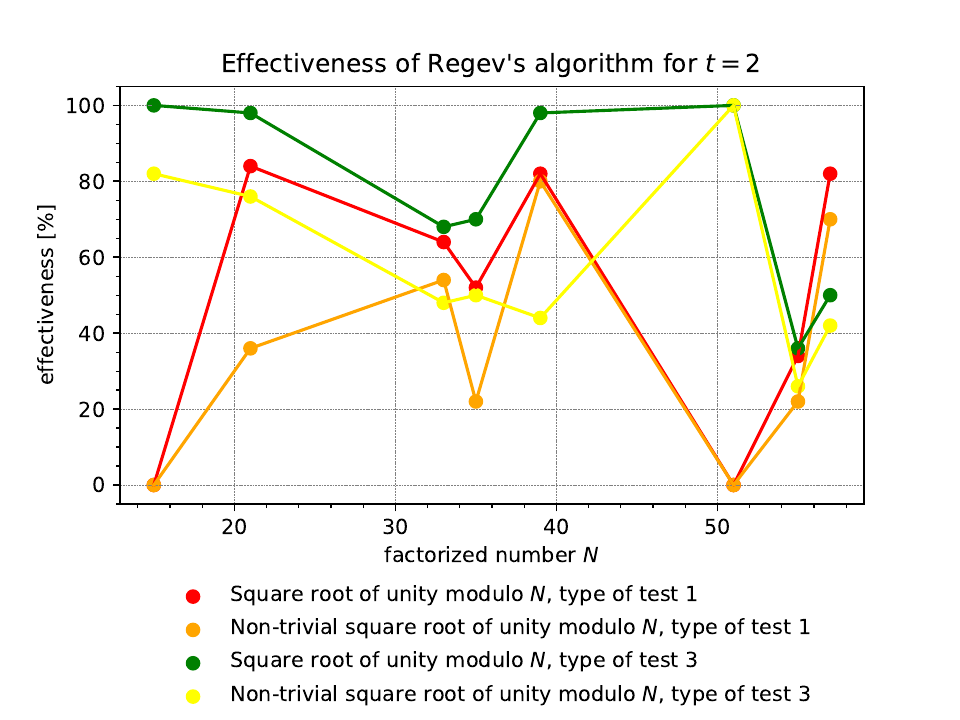}
  \caption{Effectiveness comparison for $t=2$}
  \label{fig:effectiveness_t_2}
\end{figure}

Plots for several cases were presented. Figure~\ref{fig:effectiveness_t_5} presents the results for $t=5$. We observe that finding a non-trivial square root modulo-$N$ for the third type of test exhibits a declining trend. However, for the first type of test, the results are not linear, and for $N=39$, we achieved $100\%$ effectiveness in finding a non-trivial square root modulo-$N$. Figure~\ref{fig:effectiveness_t_8} shows the results for $t=8$. For the third type of test, we again observe a declining trend. However, for the first type of test, there is a peak at $N=51$, where we achieved $100\%$ effectiveness in finding a non-trivial square root modulo-$N$. Figure~\ref{fig:effectiveness_t_14} presents results for $t=14$. For this parameter, the results for the first and third tests are similar, both showing a declining trend. However, there is a notable peak for the first test at $N=51$, with significantly better effectiveness ($52\%$) for finding a non-trivial square root modulo-$N$. Figure~\ref{fig:effectiveness_t_2} presents results for $t=2$. For this parameter, we observe that for $N=15$ and $N=21$, the effectiveness for the first test is much lower than for the third test. For higher values of $N$, the effectiveness becomes similar for both tests. 

Table~\ref{tab:effectivnes_by_t} presents, for each $N$, the parameter $t$ that yielded the highest effectiveness, together with the corresponding percentage value.
We achieved satisfactory results for all values of $N$, except for $N=55$, where the effectiveness of finding a non-trivial square root was only $32\%$. The better effectiveness of factoring $N$ than Shor's algorithm was obtained for $N \in \{21, 33, 39, 51\}$. 

\begin{table}[h!]
\caption{Effectiveness for different values of parameter $t$}
\centering
\begin{tabular}{|p{6cm}|c|c|c|c|c|c|c|}
\hline
Factorized number $N$  & \textbf{21} & \textbf{33} & \textbf{35} & \textbf{39} & \textbf{51} & \textbf{55} & \textbf{57} \\ \hline
$t$ for which obtained the highest efficiency  & 14 & 8 & 14 & 5 & 8 & 11 & 2 \\ \hline
Effectiveness of finding square root of unity modulo-$N$ $[\%]$ & 94 & 86 & 74 & 100 & 100 & 62 & 82 \\ \hline
Effectiveness of finding non-trivial square root of unity modulo-$N$ $[\%]$ & 74 & 56 & 48 & 100 & 98 & 32 & 70 \\ \hline

\end{tabular}
\label{tab:effectivnes_by_t}
\end{table}

\subsection{Performance Summary}

In Regev's algorithm, the execution time of the quantum part depends on the chosen rounding for the parameters $d$ and $\mathit{qd}$. This is because these parameters directly impact the circuit width (the number of qubits that need to be simulated) and gate complexity (the number of quantum gates in the quantum circuit). However, the faster the quantum part was executed, the lower the effectiveness of factoring the number $N$ we obtained. The execution time of Regev's algorithm was significantly greater than that of Shor's algorithm (for both the quantum and classical parts). This is because Regev's algorithm is asymptotically more efficient than Shor's algorithm, and this advantage is not apparent for smaller $N$.
In our implementation of Regev's algorithm, the parameters proposed in Kierbert's work resulted in low effectiveness in factoring the number $N$. The results we obtained were of similar to those obtained by using a random vector. For $N>35$, we observed significantly lower effectiveness compared to Shor's algorithm. In our implementation of the algorithm, due to its complexity, we did not implemented the Gaussian distribution which is correlated with parameter $t$. Therefore, we assumed that changing the value of $t$ might increase the effectiveness of our algorithm implementation.
For some values of $t$, we achieved much greater effectiveness. For some values of $N$, this even improved the effectiveness over Shor's algorithm. Table~\ref{tab:effectivnes_by_t} shows the values of $t$ for which we achieved the highest efficiency. Unfortunately, the correlation between the value of the parameter $t$ and the effectiveness of finding a non-trivial square root modulo-$N$ is not visible. However, we can conclude that the choice of parameters in the classical part has a significant impact on the algorithm's effectiveness.

\section{Conclusion}\label{concl}

This paper presents the first publicly available implementation of Regev’s quantum algorithm for quantum computers. Our primary objective was to develop and analyze the performance of the algorithm, focusing on runtime and efficiency while comparing it with Shor's algorithm. By varying parameters, we examined and visualized in graphs their influence on the execution time and effectiveness of the factorization process.

The experimental results indicate that the running time of Regev's algorithm is significantly affected by the selection of parameters $d$ and $\mathit{qd}$, which influence both the width of the circuit and the complexity of the gate. While shorter execution times led to decreased factorization effectiveness, the overall runtime of Regev’s algorithm was substantially higher than that of Shor’s algorithm.
Our implementation also revealed that the parameter selection proposed in Kierbert’s work resulted in low effectiveness in factoring $N$, with results comparable to random vector selection. However, adjusting the parameter $t$ led to improvements in effectiveness, occasionally surpassing Shor’s algorithm for specific values of $N$.
In general, our findings emphasize the importance of fine-tuning parameters to enhance the performance of Regev's algorithm. 

Future work can focus on a more comprehensive implementation, including adjustments to the Gaussian distribution, which may enhance the overall speed of the algorithm. Furthermore, more research is needed to optimize Regev’s algorithm for larger values of $N$ and real quantum hardware to fully assess its potential. Leveraging supercomputers could significantly accelerate the execution time of Regev's algorithm, enabling a more extensive evaluation of its performance across various parameter settings. This would provide a definitive testing environment for evaluating the true performance of Regev's algorithm.













\subsection*{Code Availability}

The code for the presented work, including test results, can be found in repository \url{https://github.com/Wlitkopa/regev-quantum-algorithm} (accessed on 9 July 2025).

\subsection*{Acknowledgements}


The research project was partly supported by the program ``Excellence initiative---research university'' supported by the AGH University of Krakow. This work was also supported by the EU Horizon Europe Framework Program under Grant Agreement no. 101119547 (PQ--REACT).


\begin{appendices}
\section{List of Variables}\label{secA1}
In this appendix, a brief enumeration of the most important variables used in this paper is given.

\subsection{Common Variables}

\begin{tabularx}{\linewidth}{c>{\arraybackslash}X}
$N$ & Factorized semiprime integer, product of primes $p$ and $q$.  \\
$n$ & Number of bits in the binary representation of number $N$. \\
$p$ & Prime number, factor of number $N$. \\
$q$ & Prime number, factor of number $N$. \\
$(a_1,\ldots,a_d)$ & The first $d$ squared numbers that are coprime with $N$; for convenience, the squares of primes are used. \\
$(b_1,\ldots,b_d)$ & The first $d$ numbers that are coprime with $N$; for convenience, the prime numbers are used. \\
\end{tabularx}

\subsection{Quantum Variables}

\begin{tabularx}{\linewidth}{c>{\arraybackslash}X}
$d$ & Number of dimensions in Regev's algorithm, also defines the number of quantum input registers. Its value is either equal to $\lceil \sqrt{n} \rceil$ (\texttt{ceil} version) or equal to $\lfloor \sqrt{n} \rfloor$ (\texttt{floor} version). \\
$\mathit{qd}$ & The boundary of the exponents in Regev's algorithm, also defines a width of each of the quantum input registers. Its value is either equal to $\left\lceil \frac{n}{d} + d \right\rceil$ (\texttt{ceil} version) or equal to $\left\lfloor \frac{n}{d} + d \right\rfloor$ (\texttt{floor} version). \\
\texttt{ceil\_ceil} & $d$ and $\mathit{qd}$ parameters combination when $d$ and $\mathit{qd}$ are both in \texttt{ceil} version. \\
\texttt{ceil\_floor} & $d$ and $\mathit{qd}$ parameters combination when $d$ is in \texttt{ceil} version and $\mathit{qd}$ is in \texttt{floor} version. \\
\texttt{floor\_ceil} & $d$ and $\mathit{qd}$ parameters combination when $d$ is in \texttt{floor} version and $\mathit{qd}$ is in \texttt{ceil} version. \\
\texttt{floor\_floor} & $d$ and $\mathit{qd}$ parameters combination when $d$ and $\mathit{qd}$ are both in \texttt{floor} version. \\
\end{tabularx}

\subsection{Classical Variables}

\begin{tabularx}{\linewidth}{c>{\arraybackslash}X}
$\mathcal{L}$ & A lattice containing vectors that allow computing the square root of unity modulo-$N$. \\
$\mathcal{L}_0$ & A lattice containing vectors that allow computing the trivial square root of unity modulo-$N$. \\
$\mathcal{L}'$ & A lattice used to retrieve the period vector from the vectors obtained in the quantum part. \\
$\mathcal{L}^*$ & Dual lattice of the lattice $\mathcal{L}$. \\
$B$ & A matrix whose columns form a basis of the lattice $\mathcal{L}'$. \\
$t$ & Parameter used to transform vectors returned by quantum circuit into vectors that approximates dual lattice vectors in $\mathcal{L}^*$. \\
\end{tabularx}

\end{appendices}

\bibliography{bibliography}

\end{document}